\documentclass[apj]{emulateapj}
\usepackage{apjfonts}

\begin{document}

\newcommand{\CIV}{C~{\sc iv}}
\def\sarc{$^{\prime\prime}\!\!.$}
\def\arcsec{$^{\prime\prime}\, $}
\def\arcmin{$^{\prime}$}
\def\kms{${\rm km\, s^{-1}}$}
\def\degr{$^{\circ}$}
\def\seco{$^{\rm s}\!\!.$}
\def\ls{\lower 2pt \hbox{$\;\scriptscriptstyle \buildrel<\over\sim\;$}}
\def\gs{\lower 2pt \hbox{$\;\scriptscriptstyle \buildrel>\over\sim\;$}}

\title{Dependence of the BALQSO fraction on Radio Luminosity}

\author{Francesco Shankar, Xinyu Dai, and Gregory R. Sivakoff}
\altaffiltext{1}{Department of Astronomy,
 The Ohio State University, Columbus, OH 43210,
 shankar@astronomy.ohio-state.edu, xinyu@astronomy.ohio-state.edu, sivakoff@astronomy.ohio-state.edu}


\begin{abstract}
We find that the fraction of classical Broad Absorption Line quasars
(BALQSOs) among the FIRST radio sources in the Sloan Data Release 3,
is $20.5^{+7.3}_{-5.9}$\% at the faintest radio powers detected
($L_{\rm 1.4 GHz}\sim 10^{32}\, {\rm erg\, s^{-1}}$), and rapidly
drops to $\lesssim 8$\% at $L_{\rm 1.4\, GHz}\sim 3\times 10^{33}\,
{\rm erg\, s^{-1}}$. Similarly, adopting the broader Absorption
Index (AI) definition of Trump et al. (2006) we find the fraction of
radio BALQSOs to be $44^{+8.1}_{-7.8}$\% reducing to
$23.1^{+7.3}_{-6.1}$\% at high luminosities. While the high fraction
at low radio power is consistent with the recent near-IR estimates
by Dai et al. (2008), the lower fraction at high radio powers is
intriguing and confirms previous claims based on smaller samples.
The trend is independent of the redshift range, the optical and
radio flux selection limits, or the exact definition of a radio
match. We also find that at fixed optical magnitude, the highest
bins of radio luminosity are preferentially populated by
non-BALQSOs, consistent with the overall trend. We do find, however,
that those quasars identified as AI-BALQSOs  but \emph{not} under
the classical definition, do not show a significant drop in their
fraction as a function of radio power, further supporting
independent claims for which these sources, characterized by lower
equivalent width, may represent an independent class with respect to
the classical BALQSOs. We find the balnicity index, a measure of the
absorption trough in BALQSOs, and the mean maximum wind velocity to
be roughly constant at all radio powers. We discuss several
plausible physical models which may explain the observed fast drop
in the fraction of the classical BALQSOs with increasing radio
power, \emph{although no one is entirely satisfactory}. A strictly
evolutionary model for the BALQSO and radio emission phases requires
a strong fine-tuning to work, while a simple geometric model,
although still not capable of explaining polar BALQSOs and the
paucity of FRII BALQSOs, is statistically successful in matching the
data if part of the apparent radio luminosity function is due to
beamed, non-BALQSOs.
\end{abstract}



\section{Introduction}
\label{sec|intro}

It is a main theme in Active Galactic Nuclei (AGN) studies to
describe various AGN phenomena within coherent schemes. The broad
absorption line quasars (BALQSOs) and radio quasars (RQs) are both
subsamples of quasars, where BALQSOs exhibit blue-shifted rest-frame
ultra-violet absorption troughs (e.g., Weymann et al. 1991) and RQs
are characterized by their radio emission (e.g., Matthews \& Sandage
1963; Schmidt 1963; Kellermann et al. 1989). The fractions of
BALQSOs and RQs are the subjects of many studies (e.g., Weymann et
al. 1991; Hewett \& Foltz 2003; White et al. 2007). Recently, Dai et
al. (2007, DSS hereafter), adopting the classical Balnicity Index
(BI) definition (Weymann et al. 1991) which selects as BALQSOs those
sources with absorption troughs wider than 2000$\,$\kms, obtained an
intrinsic BALQSO fraction of $(20\pm 0.2)\%$ using the BALQSOs
detected in the Two Micron All Sky Survey (2MASS) and in the Sloan
Digital Sky Survey (SDSS) Data Release 3 (DR3) BALQSO catalog of
Trump et al. (2006). DSS showed that the BALQSO fraction calculated
in the infrared is significantly higher by a factor of $\sim 2$ than
that inferred from optical samples alone. Since the infrared bands
are less affected by selection effects due to dust extinction and
absorption troughs, DSS concluded that the infrared fractions should
be much closer to the true intrinsic fractions of BALQSOs. The
intrinsic fraction from DSS is consistent with the early estimate by
Hewett \& Foltz (1993) inferred by directly correcting the
obscuration effects in the optical bands, and subsequent analysis
using similar methods by Knigge et al. (2008). In addition, such a a
high intrinsic fraction has been recently confirmed by Maddox et al.
(2008), from a deeper sample constructed from the UKIRT Infrared
Deep Sky Survey Early Data Release, which probes about three
magnitudes fainter than 2MASS. DSS also showed that an intrinsic
BALQSO fraction of 43\% is found when adopting the broader
Absorption Index (AI) classification by Trump et al. (2006) which
identifies BALQSOs as those quasars with absorption troughs which
have an equivalent width (measured between 0 and 29000 \kms) greater
than 1000$\,$\kms. The validity of the latter definition has however
been recently questioned (e.g., Ganguly et al. 2007; Knigge et al.
2008), and in this paper we will therefore discuss results adopting
both definitions.

On the other hand, recent results on the fraction of RQs in optical
samples are between 10--20\%, depending on the definition (e.g.,
Sikora et al. 2007 and references therein). In several studies, the
RQ fraction is found to depend on the redshift and optical
luminosity (e.g., Jiang et al. 2007). In general, constraining the
intrinsic fractions of BALQSOs and RQs in large, complete quasar
samples, is important for understanding the nature of these
phenomena such as their efficiency and/or black hole spin
distributions (e.g., Murray et al. 1995; Sikora et al. 2007; Shankar
et al. 2008b), and helping constrain AGN and galaxy evolutionary
models (e.g., King 2003; Croton et al. 2006; Granato et al. 2006;
Hopkins et al. 2006; Lapi et al. 2006; Shankar et al. 2006).

The fraction of BALQSOs in RQs is particularly interesting as it may
further constrain which quasar sub-samples are caused by geometrical
viewing effects and which are caused by evolutionary effects, as
well as provide additional clues on the nature and connections
between these two apparently disconnected quasar phenomena. In a
pioneer study, Stocke et al. (1992) found no evidence for luminous
radio sources among 68 BALQSOs. In a sample of 255 radio sources in
the Large Bright QSO Survey, Francis et al. (1993) noted that the
fraction of radio-moderate quasars with BAL features was
significantly higher than in the radio-quiet population. Brotherton
et al. (1998) found five BALQSOs in a complete sample of 111 QSO
candidates also detected in the NRAO VLA Sky Survey. Using the FIRST
Bright Quasar Survey, Becker et al. (2000) found 29 BALQSOs, about
14--18\% of their total sample, suggesting a frequency of BALQSOs
significantly greater than the typical $\sim$ 10\% inferred from
optically selected samples at that time. Within the small sample,
they also found that the BALQSO fraction has a complex dependence on
radio loudness. However, the trend they found was driven by the 11
low-ionization BALQSOs, whereas the 18 high-ionization BALQSOs were
slightly under-represented at high radio loudness, but at a
statistically insignificant level. With a larger sample of 25
high-ionization and 6 low-ionization BALQSOs selected from an
extension of the FIRST Bright Quasar Survey to the South Galactic
cap and to a fainter optical magnitude limit, Becker et al. (2001)
confirmed the results by Becker et al. (2000) of a steady decrease
of the high-ionization BALQSO. Using the Large Bright Quasar Survey,
Hewett and Foltz (2003) argued that with a consistent BALQSO
definition there is no inconsistency in the BALQSO fractions between
optically-selected and radio-selected quasar samples. They also
found that ``optically-bright BALQSOs are half as likely as
non-BALQSOs to be detectable $S_{1.4\rm GHz} \gtrsim 1$~mJy as radio
sources''. In this paper, we combined the SDSS-DR3 BALQSO sample
(Trump et al. 2006), containing more than 4,000 BALQSOs, with the
FIRST survey data (Becker et al. 1995) to study the fraction of
BALQSOs in the radio detected sample.

\section{The Data}
\label{sec|Data}

The parent sample for the Trump et al. (2006) BALQSO catalog is the
SDSS DR3 quasar catalog (Schneider et al. 2005). Data were taken in
five broad optical bands ($ugriz$) over about 10,000 deg$^2$ of the
high Galactic latitude sky. The majority of quasars were selected
for spectroscopic followup by SDSS based on their optical
photometry. In particular, most quasar candidates were selected by
their location in the low-redshift ($z \lesssim 3$) $ugri$ color
cube with its $i$-magnitude limit of 19.1. A second higher redshift
$griz$ color cube was also used with a fainter $i$-magnitude limit
of 20.2. Following DSS, we focus on redshift range of $1.7 \le z \le
4.38$, where BALQSOs are identified by \CIV\ absorption in the SDSS
spectra. In the following we present results using both the AI and
BI definitions for BALQSOs.

The DR3 quasar catalog by Schneider et al. (2005) also provides the
radio flux for those sources which have a counterpart within
2\arcsec in the FIRST catalog (Becker et al. 1995). However, we use
the integrated flux density observed at 1.4 GHz as listed in the
FIRST catalog to calculate the specific luminosity emitted at 1.4
GHz, $L_{1.4 {\rm \, GHz }}$, via the relation
\begin{equation}
L_{1.4 {\rm GHz}} = \frac{f_{\rm int} \times 10^{-26}}
{(1+z)^{1-\alpha}}4\pi D_L^2 {\rm \, erg \,s^{-1} \, Hz}^{-1}\, ,
\label{eq|L14GHz}
\end{equation}
where $f_{\rm int}$ is the integrated flux density in mJy, $\alpha$
is the spectral index (i.e., $f(\nu) \propto \nu^{-\alpha}$), and
$D_L$ is the luminosity distance in cm as calculated from the
redshift setting $\Omega_m=0.3$, $\Omega_{\rm \Lambda}=0.7$ and
$h=0.7$. We assume $\alpha=0.7$, which is intermediate of the
typical range for radio jets (Bridle \& Perley 1984). Our results do
not qualitatively change if we assume either $\alpha=0.5$ or
$\alpha=1.0$.

According to Schneider et al. (2005), while only a small minority of
FIRST-SDSS matches are chance superpositions, a significant fraction
of the DR3 sources are extended radio sources. Therefore the FIRST
catalog position for the latter may differ by more than 2\arcsec
from the optical position. To perform a comprehensive study of the
BALQSO radio fraction, we built a full FIRST-SDSS cross-correlation
catalog, containing all the detected radio components within
30\arcsec of a optical quasar. As our reference, following Schneider
et al. (2005), we primarily present results for the FIRST-SDSS
catalog with radio counterparts defined within 2\arcsec, which
contains a total of 652 radio sources, out of which 78 are BALQSOs,
about three times the sample used by Becker et al. (2001) for
BALQSOs. We will discuss results obtained by enlarging the radio
matches to 5\arcsec. In addition, many of the optical sources in
SDSS are associated to more than one radio component in FIRST within
30\arcsec, as expected if these sources are extended with jets
and/or lobes separated from the central source. We will therefore
discuss results after the inclusion of BALQSO sources characterized
by multiple radio components, for which we use the sum of the flux
densities as a proxy for the total radio luminosity of the source. 

Finally, we also stress that FIRST efficiently identifies radio
matches to optically-selected quasars. By cross-correlating SDSS
with the large radio NRAO VLA Sky Survey (Condon et al. 1998), Jiang
et al. (2007) found in fact that only $\sim$ 6\% of the matched
quasars were not detected by FIRST. Most of these sources have
offsets more than 5\arcsec and are uniformly distributed in
luminosity thus not altering the main conclusions of this paper.

\section{Results}
\label{sec|Results}

\subsection{The Radio fraction of BALQSOs}
\label{subsec|fbalradio}

\begin{figure*}[ht!]
\epsscale{1.}
\plotone{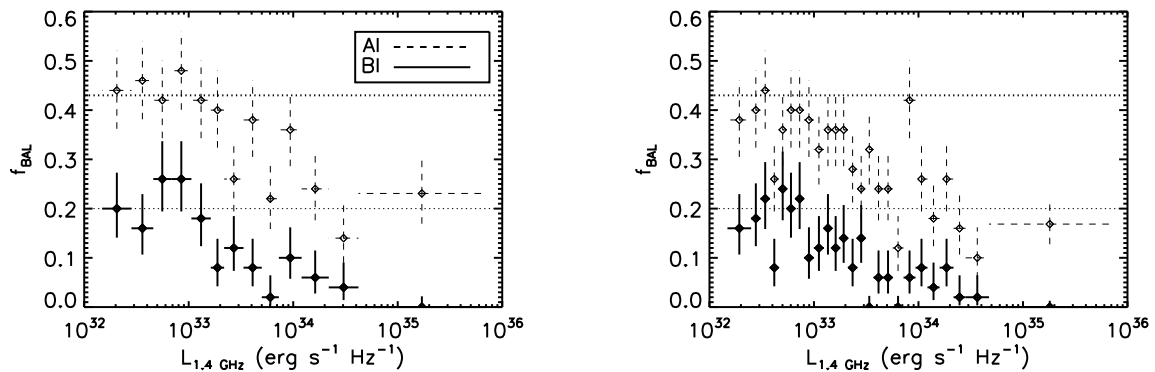} \small{\caption{\emph{Left panel}: fraction of
BALQSOs within the radio sample of SDSS quasars with optical
luminosity brighter than $m_i=19.1$ as a function of radio
luminosity and with radio counterpart within 2\arcsec of the optical
source; the \emph{filled symbols} with solid lines show the sample
of BALQSOs defined under the BI definition, while the \emph{open
symbols} with dashed lines show the sample of BALQSOs sample
compiled following the AI definition; the \emph{lower dotted} and
\emph{upper dotted} lines refer to the ``intrinsic fractions'' of
20\% and 43\% of BALQSOs found in 2MASS by Dai et al. (2007),
respectively. \emph{Right panel}: fraction of BALQSOs within the
\emph{full} SDSS sample of RQs; here all lines and symbols are as in
the left panel; the statistical significance of the drop in the
radio-BALQSOs fraction towards high radio powers does not depend on
the optical magnitude limit.}\label{fig|fBAL}}
\end{figure*}

The solid crosses in the left panel of Figure~\ref{fig|fBAL} show
the fraction
\begin{equation}
f_{\rm BAL}=\frac{N_{\rm BAL,RADIO}}{N_{\rm RADIO}}\, ,
    \label{eq|fbal}
\end{equation}
as a function of radio luminosity, where $N_{\rm RADIO}$ is the
total number of RQs detected in SDSS and $N_{\rm BAL,RADIO}$ is the
number of RQs which are also classified as BALQSOs. For this plot we
only used the ``complete'' SDSS optical sample with apparent optical
magnitude $m_i\le 19.1$. The bins in $L_{1.4 {\rm \, GHz }}$ were
created adaptively starting from low radio luminosities by requiring
a statistically significant sub-sample of 50 quasars in each bin of
luminosity. In each bin we used the binomial theorem to estimate
$f_{\rm BAL}$ along with its $\pm 1\sigma$ confidence levels (e.g.,
Gehrels et al. 1986).

The first important result of our analysis is the high fraction of
classical (BI) BALQSOs detected at low radio powers. The BALQSO
fraction of $20^{+7.3}_{-5.9}$\% at the faintest radio powers of
$L_{1.4 {\rm \, GHz }}\lesssim 10^{33}\, {\rm erg\, s^{-1}}$ is in
strikingly good agreement with the 20\% classical BALQSO occurrence
inferred by DSS in the 2MASS-SDSS sample (shown with horizontal,
thin-dotted line in Figures~\ref{fig|fBAL} and \ref{fig|fBALfull}).
This result further supports and extends the key result discussed in
DSS that the detected BALQSO fraction should be closer to the
intrinsic one at longer wavelengths, which are gradually less
affected by absorption troughs and dust extinction.

The second striking result of our study is that $f_{\rm BAL}$ drops
rapidly with increasing radio power to $8.0^{+5.9}_{-3.8}$\% at
$L_{1.4 {\rm \, GHz }}\gtrsim 2\times 10^{33}\, {\rm erg\, s^{-1}}$.
This behavior strongly supports some close connection between radio
and BALQSO phenomena and, as discussed below, poses serious
challenges to simple evolutionary models.

The open symbols with dashed crosses represent the fraction $f_{\rm
BAL}$ obtained including all AI-selected BALQSOs. As expected, we
get a much larger fraction of BALQSOs within our sample of
$44^{+8.1}_{-7.8}$\%, in very good agreement with the findings by
DSS who found a fraction of 43\% AI BALQSOs in 2MASS (thick-dotted
line). Even in this case, $f_{\rm BAL}$ presents a very similar
behavior as a function of radio luminosity and that we will discuss
in \S~\ref{subsec|robust}.

\subsection{Robustness of Results}
\label{subsec|robust}

Our results are derived by cross-correlating optical quasars, RQs,
and BALQSOs. Therefore it is natural to expect that several issues
may affect our conclusions. In this section we carefully analyze the
most significant sources of bias which can enter our computations
and find that none of these seems to be responsible for the observed
$f_{\rm BAL}-L_{1.4 {\rm \, GHz }}$ behavior.

We first tested for biases from redshift or luminosity selection
effects. When repeating our analysis using only the sources in two
redshift bins $1.7\lesssim z \lesssim 2.5$ and, separately,
$2.5\lesssim z \lesssim 4.38$, we find very similar trends for the
$f_{\rm BAL}$-$L_{1.4 {\rm \, GHz }}$ relation.

The completeness of the Schneider et al. (2005) quasar sample has
been studied in detail by Vanden Berk et al. (2005), who estimate a
total completeness of 89.3\% down to $m_i=19.1$. We do not expect
the behavior of $f_{\rm BAL}$ to be severely influenced by these
mild incompleteness effects. As extensively discussed (e.g.,
Cirasuolo et al. 2003; Shankar et al. 2008b, and references
therein), the radio and optical luminosities are correlated with a
large observed scatter. Therefore, incompleteness effects in the
optical sample will, on average, induce similar effects in the
numerator and denominator of equation~(\ref{eq|fbal}), given the
extended distribution of optical sources at fixed radio luminosity.
As an example of the effect of different optical luminosity
thresholds on the fraction of BALQSOs, we show the resulting $f_{\rm
BAL}$ with no optical magnitude cut in the right panel of
Figure~\ref{fig|fBAL}. This highly incomplete sample in the optical
still shows the exact same drop of at least a factor of $\sim 4$ in
$f_{\rm BAL}$.

\begin{figure}[ht!]
\epsscale{1.}
\plotone{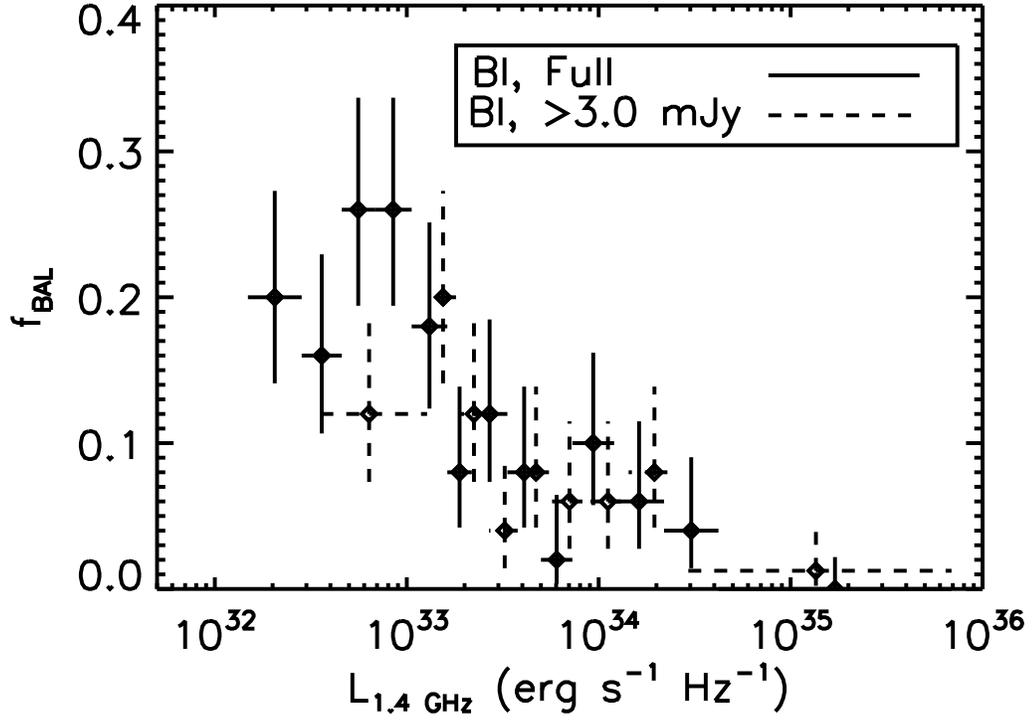} \caption{Dependence of $f_{\rm BAL}$ on radio
luminosity for samples defined to have different flux density limits
$f_{\rm int}$ in FIRST and corresponding different completeness
levels. The drop in $f_{\rm BAL}$ is always present.}
\label{fig|fractionFlim}
\end{figure}

On the other hand, our sample may suffer from selection effects in
FIRST. At 3 mJy the completeness of FIRST is about $93$\%,
decreasing to 75\% at 1.5 mJy, and 55\% at 1.1 mJy (see Figure 1 in
Jiang et al. 2007). To check whether our results were affected by
this incompleteness, in Figure~\ref{fig|fractionFlim} we compare the
results shown in the left panel of Figure~\ref{fig|fBAL} obtained
from the full sample (solid crosses), with the sample defined by
only those radio sources with $f_{\rm int}>3$ mJy (dashed crosses).
This cut in radio flux density ensures the best completeness as
possible from FIRST. It is evident that the strong observed drop in
$f_{\rm BAL}$ remains unchanged.

\begin{figure*}[ht!]
\epsscale{1.}
\plotone{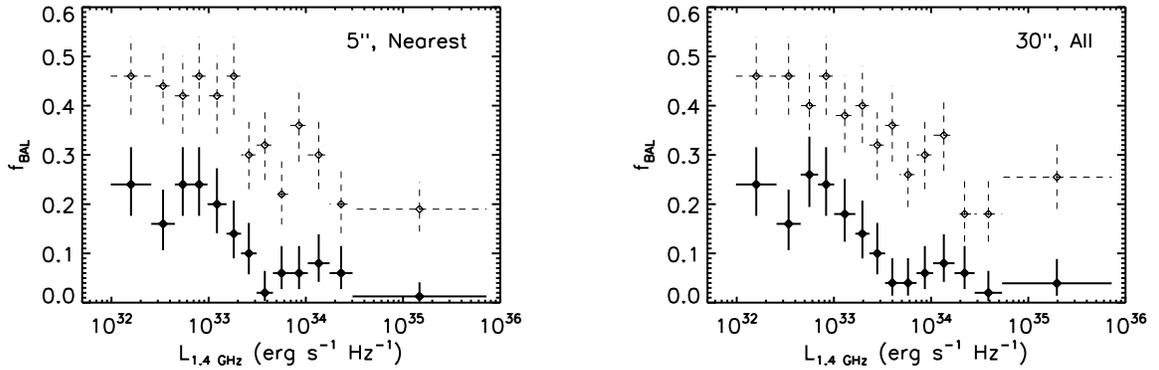} \small{\caption{\emph{Left panel}: fraction as a
function of radio power of radio-BALQSOs with an optical magnitude
brighter than $m_i=19.1$ with a single radio spot within 5\arcsec of
the optical source; if multiple radio components are contained
within this aperture, then only the closest source is considered.
\emph{Right panel}: fraction of radio-BALQSOs for sources which lie
within 30\arcsec of the optical source; if multiple radio components
are present within this aperture, then they are treated as a single
source with radio power given by the sum of the integrated flux
densities of each component. In both panels all symbols and lines
have same meaning as in Figure~\ref{fig|fBAL}.}\label{fig|fBALfull}}
\end{figure*}

As anticipated in \S~\ref{sec|Data}, the 2\arcsec cross-correlation
sample ensures a \emph{secure} identification of the core radio
power associated to an optical quasar. However, to check for any
bias in our results due to incomplete radio identifications in the
SDSS DR3 quasar catalog, we recomputed $f_{\rm BAL}$ using the full
sample built by matching SDSS and FIRST. The left panel of
Figure~\ref{fig|fBALfull} shows the fraction of BALQSOs with the
radio counterpart defined by the nearest source in FIRST within
5\arcsec. The right panel of Figure~\ref{fig|fBALfull} shows instead
the fraction of BALQSOs with at least one radio counterpart within
30\arcsec of the optical source. In both cases we still find $f_{\rm
BAL}$ to drop with increasing radio power. These results are
actually expected, as a closer inspection revealed that the overall
sample of radio-optical sources increases by just $\sim 4$\% when
including sources within 5\arcsec, out of which only 3 extra BALQSOs
are found with respect to our reference sample. The increase is even
milder ($\sim 3$\%) when the 30\arcsec sources are included, with
only 2 extra BALQSOs included in the sample. The paucity of extended
radio sources in our sample is in agreement with Gregg et al. (2006)
who found only six quasars from the DR3 and FIRST Bright Quasar
Survey that exhibit Fanaroff-Riley II-type (FRII) morphologies and
BAL properties, with a BI index strongly decreasing with increasing
radio loudness.

\begin{figure}[ht!]
\epsscale{1.}
\plotone{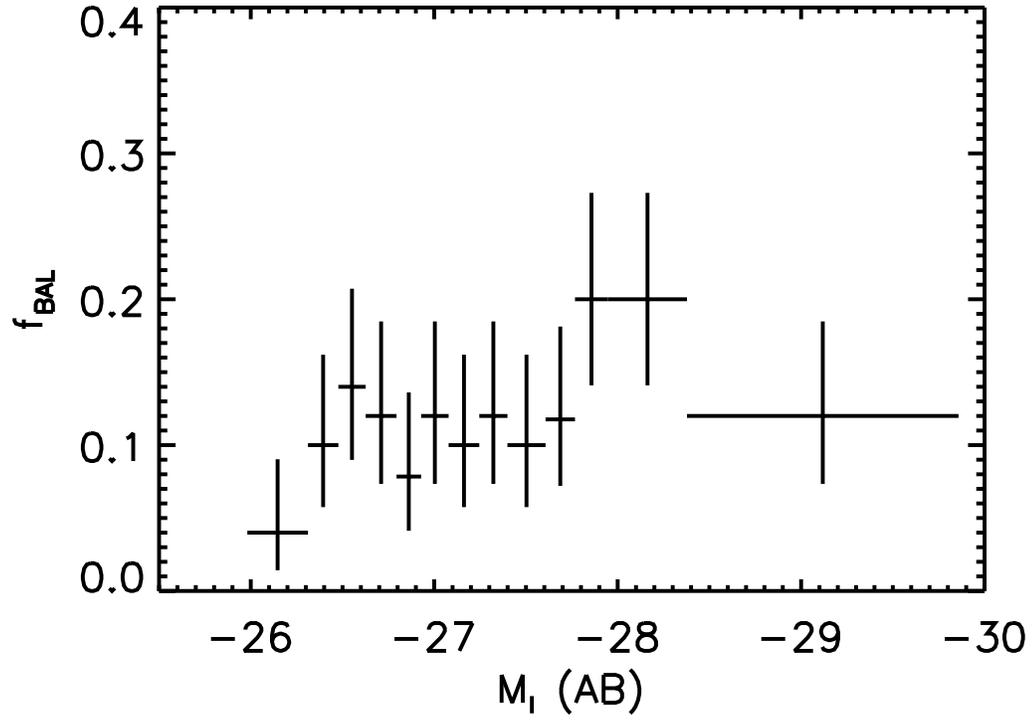} \caption{Fraction of radio-BALQSOs as a function
of AB optical magnitude. $f_{\rm BAL}$ is about constant if not
increasing at higher optical luminosities, which does not supports
any direct obvious link in the drop of $f_{\rm BAL}$ with increasing
radio power. Note that the true fractions in the optical should be
higher once corrected for absorption and extinction (Dai et al.
2008).} \label{fig|fbalMopt}
\end{figure}

In Figure~\ref{fig|fbalMopt} we show $f_{\rm BAL}$ as a function of
absolute optical magnitude. We do not find significant evidence for
any sharp trend of $f_{\rm BAL}$ as a function of $M_i$, confirming
the results by DSS for the 2MASS sample. This implies that different
optical magnitude cuts do not affect $f_{\rm BAL}$ on one hand, and,
on the other, implies that the drop in $f_{\rm BAL}$ is mainly a
``radio phenomenon'', not linked with the quasar bolometric
luminosity. Note that we did not apply any correction in
Figure~\ref{fig|fbalMopt} for extinction as was done in DSS,
therefore the resulting fractions in the optical are lower than the
intrinsic ones (see DSS for details).

\begin{figure*}[ht!]
\epsscale{1.}
\plotone{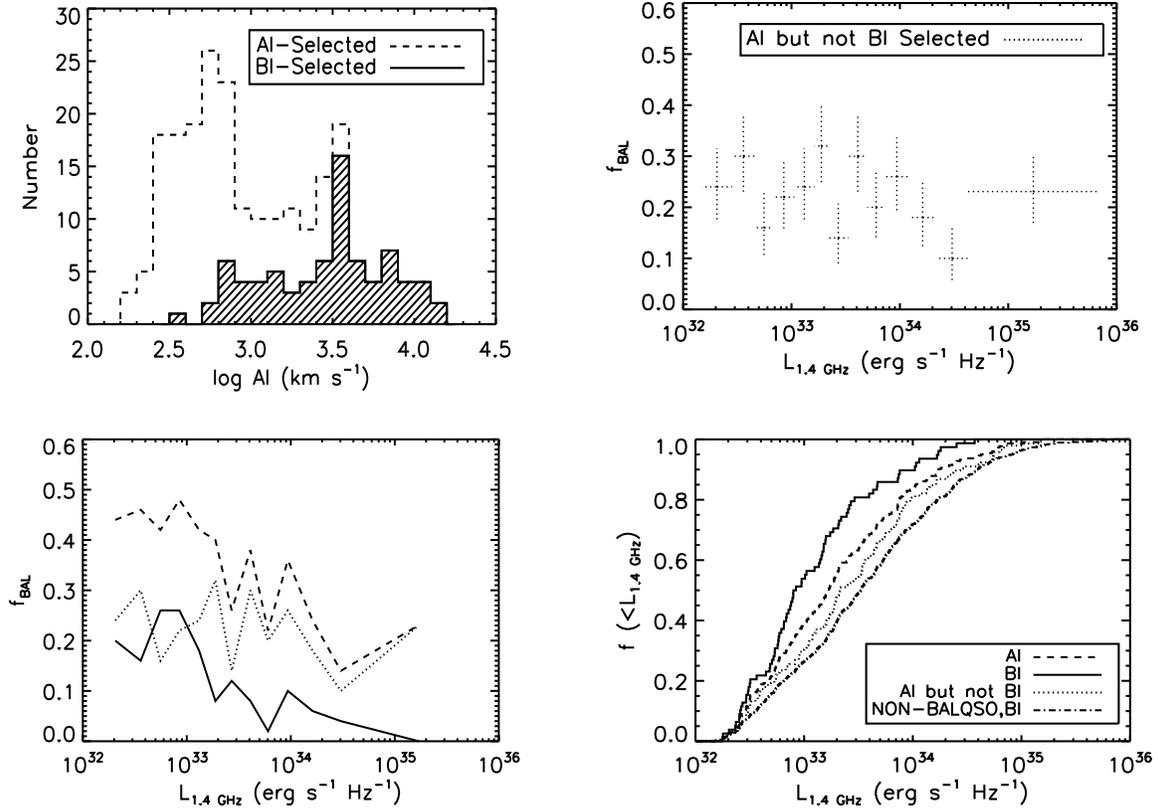} \caption{\emph{Upper left panel}: histogram of
all BALQSOs defined with the AI definition; the striped histogram
refers to the fraction of the sources which are identified as
BALQSOs also under the BI definition. \emph{Upper right panel}:
fraction of sources defined as BALQSOs with the AI definition but
not with the BI one as a function of radio power; the drop in
$f_{\rm BAL}$ with increasing radio power is much less evident for
this subclass of sources. \emph{Lower left panel}: the
\emph{dashed}, \emph{dotted} and \emph{solid} lines show the
fractions of radio-BALQSOs defined with the AI definition only, with
AI but not BI, and with the BI definition, respectively, as a
function of radio power. \emph{Lower right panel}: cumulative
distributions for the different subclasses of BAL and non-BALQSOs,
as labeled, as a function of radio power. The probability that AI
but not BI selected BALQSOs are drawn from the same distribution as
non-BALQSOs is $P=0.26$.} \label{fig|AIvsBI}
\end{figure*}

The AI sample by Trump et al. (2006) may suffer from a significant
contamination of false positives, which the close inspection by
Ganguly et al. (2007) found to be about 30\% (see also Knigge et al.
2008). To investigate the impact these findings may have on our
results we have compared the sample of sources cataloged as BALQSOs
using the AI definition, with the sample of BALQSO defined using the
classical BI index. The distribution of AI-BALQSOs with radio
counterparts as a function of the AI index (open histogram) is
compared in the upper left panel of Figure~\ref{fig|AIvsBI} with the
distribution of the sources which are also identified as BI-BALQSOs
(filled histogram). We find the latter sample strongly peaked at
higher AI values, while just the opposite is true for most of the
AI-BALQSOs. These findings are fully consistent with the recent
results by Knigge et al. (2008) who claim for a well distinct
separation for the two BALQSO samples. The upper right panel of
Figure~\ref{fig|AIvsBI} plots the fraction of AI- but \emph{not}
BI-identified BALQSOs as a function of radio power: the drop in
$f_{\rm BAL}$ with increasing radio power is much less evident for
this subclass of sources compared to BI-BALQSOs. These results
suggest that AI-BALQSOs are a different class of quasars with
respect to the BI-BALQSOs. In the lower right panel of
Figure~\ref{fig|AIvsBI} the dashed, dotted and solid lines show the
fractions of AI-BALQSOs, AI- but not BI-BALQSOs, and BI-BALQSOs,
respectively. This plot confirms that the apparent drop in the
AI-BALQSOs with radio power is driven by the behavior in BI-BALQSOs.
The lower right panel of Figure~\ref{fig|AIvsBI} shows the
cumulative distributions for the different subclasses of BAL and
non-BALQSOs, as labeled. We find that the probability that AI but
not BI selected BALQSOs are drawn from the same distribution as
non-BALQSOs is $P=0.26$, while the probability for being drawn from
any other distribution drops by several orders of magnitude.

\begin{figure*}[ht!]
\epsscale{.6}
\plotone{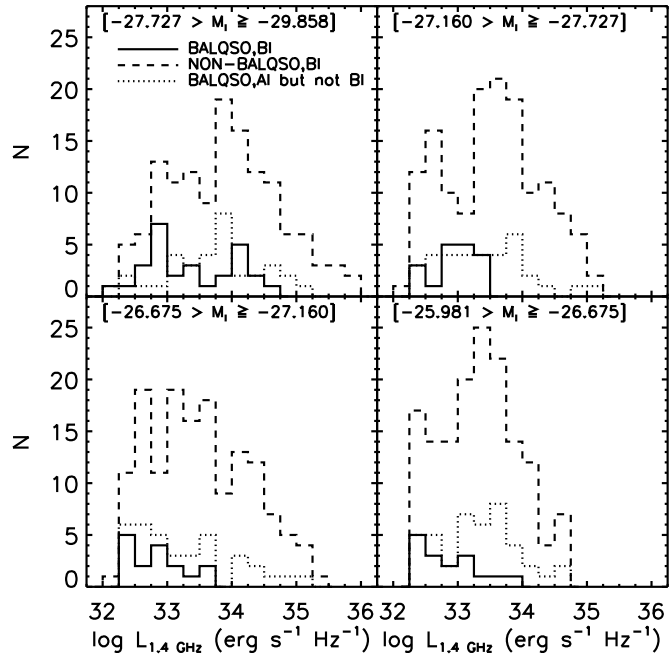} \caption{The \emph{dashed}, \emph{dotted}, and
\emph{solid} lines show, respectively, the histograms of the
distributions of BI-BALQSOs, AI but not BI BALQSOs, and non-BALQSOs
as a function of radio luminosity in different bins of optical
luminosity, as labeled. Unlike the BI BALQSOs, the non-BI BALQSOs
have a greater tendency to form a prolonged tail of very luminous,
potentially beamed, radio sources.} \label{fig|cumul}
\end{figure*}

\begin{figure}[ht!]
\epsscale{1.}
\plotone{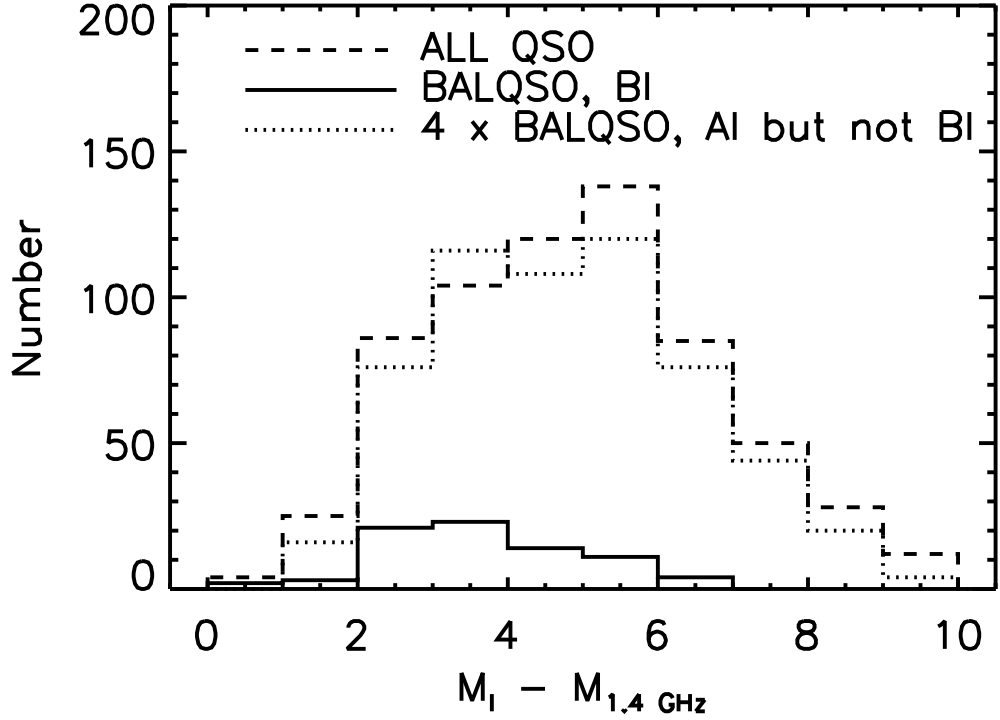} \caption{The \emph{dashed}, \emph{dotted}, and
\emph{solid} lines show, respectively, the cumulative histograms of
the distributions of BI-BALQSOs, AI but not BI BALQSOs, and
non-BALQSOs as a function of color. This plot shows that on average
at fixed optical luminosity the non-BI BALQSOs have a tendency to be
more radio luminous than BI BALQSOs.} \label{fig|cumulTOT}
\end{figure}

In Figure~\ref{fig|cumul} we show the distributions of all
non-BALQSOs (dashed lines), BI-BALQSOs (solid lines), and AI- but
not BI-BALQSOs (dotted lines) as a function of radio luminosity for
different bins of absolute optical magnitude, as labeled. These bins
were chosen to provide an equal number of BALQSOs in each panel. It
is apparent that compared to BI-BALQSOs, both the AI- but not
BI-BALQSOs and the non-BALQSO distributions have a greater tendency
to populate significantly higher, potentially beamed, radio
luminosity bins. This provides further evidence that the trends in
$f_{\rm BAL}$ are not a mere consequence of random effects in the
radio power distributions of optical quasars. The dashed, dotted,
and solid lines in Figure~\ref{fig|cumulTOT} show, respectively, the
histograms of the distributions of BI-BALQSOs, AI- but not BI-
BALQSOs, and non-BALQSOs as a function of color defined as the
difference between absolute $i$-band and AB radio magnitudes listed
in Schneider et al. (2005). This plot shows that on average AI- but
not BI- BALQSOs, and non-BALQSOs have very similar optical-to-radio
distributions while BI-BALQSOs are clearly offset populating lower
radio luminosity bins at fixed optical magnitude.

Overall our results show that the classical BALQSO fraction in the
radio is as high as inferred from the near-IR and rapidly drops from
$\sim 20\%$ to $\lesssim 4\%$ with increasing radio power. This
clear trend confirms previous claims by Becker et al. (2001), taking
advantage of a sample of $\sim 3$ higher. We conclude that the
behavior of $f_{\rm BAL}$ as a function of radio power seems to be
an intrinsic, real feature of the radio, BI-BALQSO population, not
linked to any significant selection effect. In the following, if not
alternatively specified, we will always refer to BALQSOs as those
defined through the classical BI definition.

\section{Explaining the observed trend}
\label{sec|models}

Models aimed at explaining the BALQSO and/or radio fractions, can be
broadly divided into two classes: one in which each quasar phase is
distinct and characterized by the internal physical \emph{evolution}
undergone by the system, and the other in which all these phenomena
co-exist at all times but different source \emph{orientations}
modify the observed balance of one process with respect to another.

\subsection{The Evolutionary Model}
\label{subsec|evolutionmodel}

\begin{figure}[ht!]
\epsscale{0.8}
\plotone{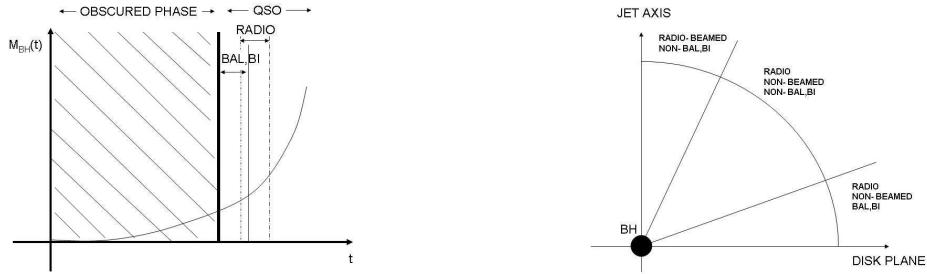} \caption{\emph{Left panel}: schematic diagram of
a pure ``evolutionary'' model. The mass of the central black hole
starts growing in a heavily obscured phase. When it becomes
sufficiently luminous to blow away the surrounding medium, it can
then be detected as a BALQSO and then finally as an optical,
dust-free quasar. \emph{Right panel}: schematic representation of a
simple ``orientation'' model for RQs and BALQSOs. The radio phase in
both schemes is always considered as a brief period within the
optical phase.} \label{fig|cartoon}
\end{figure}

The results discussed in \S~\ref{sec|Results} seem to pose a serious
challenge to ``evolutionary'' models for BALQSOs. The simple sketch
shown in the left panel of Figure~\ref{fig|cartoon}, represents the
basic picture for the evolution of a black hole within its host
galaxy (e.g., Kawakatu et al. 2003). The mass of the central black
hole, shown as a solid line in the left panel of
Figure~\ref{fig|cartoon}, exponentially grows through gas accretion
from the surrounding medium. Initially the whole galaxy, or at least
its central region, is buried under optically-thick layers of dust
which allow the AGN to be detectable only in those bands that suffer
less from obscuration, such as in X-rays (e.g., Ueda et al. 2003).
When the black hole becomes sufficiently luminous, it injects enough
energy and momentum (e.g., Fabian 1999) into the surrounding medium
to disperse the obscuring material. During this blowout phase, the
system will be detected as a BALQSO featuring the signatures of
strong winds in their spectra, while at later stages it will be
observed as an optical, dust-free quasar.

It is still unclear what triggers the radio activity of an AGN
(e.g., Sikora et al. 2007), so it is difficult to predict when an
optical quasar should become radio-loud (e.g., Blundell \& Kuncic
2007; Shankar et al. 2008b). Nevertheless, in pure evolutionary
models the radio phase is usually viewed as a brief period within
the optical phase as the incidence of RQs in optical samples is
rather small, ($\sim 15-20\%$, e.g., Jiang et al. 2007; see also
\S~\ref{sec|intro}). In this model, our results on $f_{\rm BAL}$
indicate that the radio duty cycle must overlap the boundary between
the BALQSO and optical phases, because a significant fraction of
optical quasars also emit in radio but are not BALQSOs, and vice
versa.

In a pure evolutionary model our result that $f_{\rm BAL}\sim 20\%$
constrains the BALQSO and non-BALQSO phases to last for $\sim$ 20\%
and 80\%, respectively, within the brief radio phase, which is about
15--20\% of the total optical quasar lifetime (see left panel of
Figure~\ref{fig|cartoon}). The consistency between the optical and
radio BALQSO fractions imposes a coincidence problem, which could be
otherwise naturally interpreted in orientation models
(\S~\ref{subsec|beamingmodel}). Additionally, the drop in $f_{\rm
BAL}$ at high radio powers is difficult to reproduce simply in pure
evolutionary models, and thus such models would need to be
fine-tuned to explain the drop. One simple explanation for such a
trend may be an increasing feedback efficiency with quasar
luminosity (e.g., Kelly et al. 2008). The more luminous sources
would then be faster in removing the surrounding obscuring medium,
shortening the BALQSO phase and consequently inducing a decrease in
$f_{\rm BAL}$. However, this model is at variance with the results
in Figure~\ref{fig|fbalMopt} and by DSS which show that the BALQSO
fraction is constant with optical luminosity.

\begin{figure}[ht!]
\epsscale{1.0}
\plotone{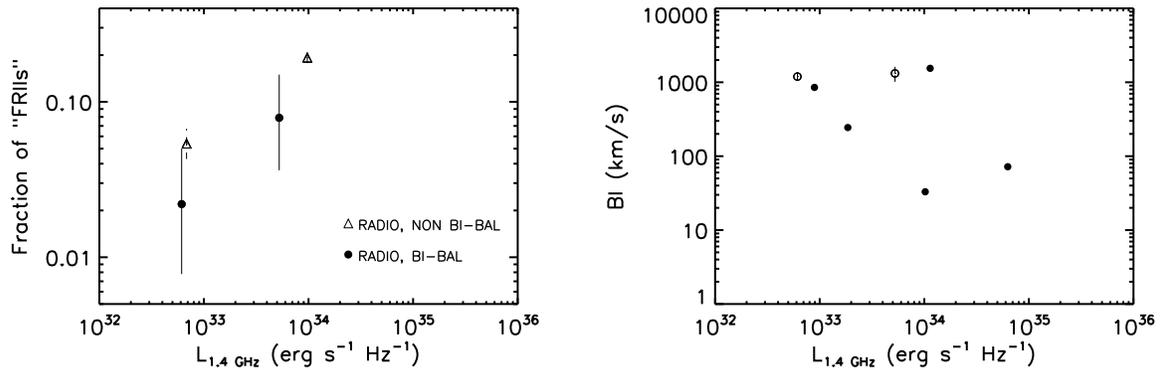} \caption{\emph{Left panel}: fraction of
FRII-selected sources among the BI-selected BALQSOs (\emph{filled
circles}) and non-BALQSOs (\emph{open triangles}); both samples have
been divided into two subgroups with radio luminosities below and
above $L_{\rm 1.4 GHz}=2\times 10^{33}\, {\rm erg\, s^{-1}\,
Hz^{-1}}$; we display the median luminosity of each subsample.
\emph{Right panel}: absorption index for non-FRII radio sources
(\emph{open circles}) and radio-BI selected FRII sources
(\emph{filled circles}) as a function of radio power. For the more
numerous non-FRIIs, we display the mean BI values and associated
uncertainties following Hoaglin et al. (1983; see
\S~\ref{subsec|energyexchange}).} \label{fig|Gregg}
\end{figure}

Gregg et al. (2006) found an absence of BALQSO FRII sources and a
drop in the BI index with increasing radio loudness of the sources.
They interpreted their results within an evolutionary scheme where a
BALQSO occurs in the early stage of a quasar during its emergence
from a thick shroud of surrounding dust (see left panel of
Figure~\ref{fig|cartoon}). In this model FRIIs are only seen after
most of the obscuring shroud has been removed. Following Gregg et
al. (2006) we further probe the constraints for evolutionary models
by addressing the coexistence of FRII/BAL sources within our sample.
In Figure~\ref{fig|Gregg} the open triangles show the fraction of
FRII-selected sources\footnote{Here we simply call ''FRIIs'' all
those sources in the sample with multiple radio components within
30\arcsec.} within the whole radio sample, while the filled circles
refer to the fraction of FRII sources within the BI-selected radio
sample. Both samples have been divided into two subgroups with radio
luminosities below and above $L_{\rm 1.4 GHz}=2\times 10^{33}\, {\rm
erg\, s^{-1}\, Hz^{-1}}$. Note that the contamination from random
interloping radio sources in the identification of FRII sources is
small. The average number of interlopers within 30\arcsec of any
position is only 0.02 (Becker et al. 1995), i.e., a cumulative
number of just about 3 sources are expected to be misidentified as
FRIIs. It can be seen that in both BI and not BI-selected samples,
the fraction of FRII increases with radio power, mirroring the fact
that FRIIs are intrinsically more luminous radio sources (e.g., De
Zotti et al. 2005). We do find some evidence for FRIIs to be less
associated to BALQSOs within the BI-selected sources than in the
overall sample at fixed radio power, although the difference is only
marginally significant. Gregg et al. (2006) claim a drop by a about
an order of magnitude in the fraction of FRIIs in BAL-radio samples,
while we find a mean drop of $\sim 2.4$. Also, we do not find strong
evidence for the Gregg et al. (2006) anti-correlation between BI
index and radio loudness. A Pearson's \emph{r}-test yields a
correlation parameter of $r=-0.36$, with a significance of $P=0.55$;
however, the small number of FRII BALQSOs limits this analysis. The
BI index of FRIIs in our BAL-radio sample does not show any clear
trend with radio power, as shown with filled circles in the right
panel of Figure~\ref{fig|Gregg}. For comparison, in the same Figure
we also show with open symbols the mean BIs for the low and high
radio power subsamples of non-FRIIs, which show a rather flat
behavior with increasing radio power, in agreement with the overall
sample (see \S~\ref{subsec|energyexchange} and Figure~\ref{fig|AI}).

It also may well be true that the BALQSOs we are referring to here
are only a partial representation of the ``outflowing'' AGNs. Many
more systems may be detected during initial and final phases of the
blowout and therefore may be characterized by lower velocities.
Ganguly \& Brotherton (2008) have shown that including Narrow
($\lesssim 800\, $\kms) and Associated absorption systems, the
latter being characterized by narrow absorption-lines near the
redshift of the quasar, the fraction of outflowing AGNs rises to
about 60\%, independent of optical luminosity. The $f_{\rm BAL}$
radio dependence for the cumulative sample may then be different
then what is inferred here. Several works have shown that a
significant fraction of RQs show associated absorption and high
radio power sources may preferentially exhibit narrower absorption
features (e.g., Weymann et al. 1979; Anderson et al. 1987; Ganguly
et al. 2001; Vestergaard 2003). Such a hypothesis needs to be tested
with samples with identified Associated Absorption systems, such as
those by Ganguly et al. (2007). Nevertheless, the drop in $f_{\rm
BAL}$ for BI-BALQSOs will still need to be explained.

\subsection{The Orientation-Beaming Model}
\label{subsec|beamingmodel}

In this section we show that a simple orientation effect can fully
explain the puzzling trend in the $f_{\rm BAL}$-$L_{1.4 {\rm \, GHz
}}$ distribution, although even this model is not entirely
satisfactory, as discussed at the end of the section. This model
assumes that some fraction of RQs are relativistically beamed
towards the observer and are boosted to higher radio luminosities.
Since the radio luminosity function is steep, beamed radio sources
are a disproportionate fraction of the bright sources. According to
popular disk wind models (e.g., Murray et al. 1995, Proga et al.
2000), radio BALQSOs are instead on average non-beamed sources,
being preferentially viewed close to the plane of the accretion disk
surrounding the central black hole. It is then reasonable to expect
that if BALQSOs are a fixed fraction of the \emph{intrinsic} radio
luminosity function, then the observed occurrence of BALQSOs must
decrease at high radio powers due to the apparent increase of
$N_{\rm RADIO}$ in Eq.~(\ref{eq|fbal}) towards high luminosities due
to beaming.

\begin{figure*}[ht!]
\epsscale{1.0}
\plotone{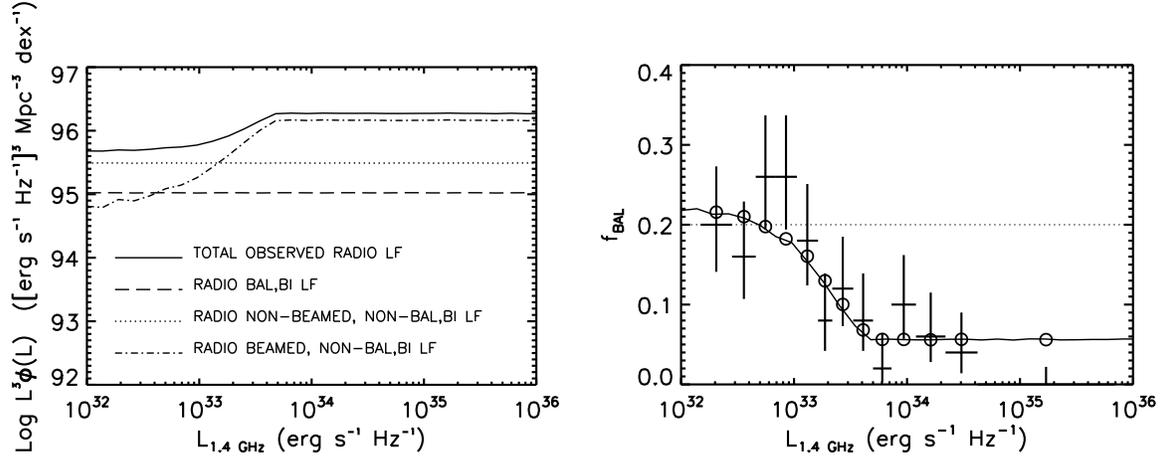} \caption{\emph{Left panel}: radio quasar
luminosity functions plotted in units of $L^3\Phi(L)$ to enhance the
differences between the lines; the \emph{long-dashed} line is the
luminosity function of RQs detected as BI BALQSOs, normalized to be
20\% the one of the intrinsic total radio quasar luminosity
function, which we have assumed to have a slope of $-3.0$; the
\emph{dotted} line is the fraction of $\sim 59\%$ of radio
non-beamed, non-BI BALQSOs, while the \emph{dot-dashed} line is the
fraction of $\sim 21\%$ of beamed sources (see text for details);
the \emph{solid} line is the total \emph{apparent} luminosity
function. \emph{Right panel}: the \emph{solid} line is the expected
fraction of BALQSOs as a function of radio luminosity derived as the
ratio of the \emph{dashed} and \emph{solid} lines in the left panel;
the \emph{open circles} show the expected fraction averaged over the
same radio luminosity bin as the data; all other symbols and lines
are as in the left panel of Figure~\ref{fig|fBAL}.}
\label{fig|simul}
\end{figure*}

We quantitatively explore this basic idea by building a simple
``unification'' model (e.g., Urry \& Padovani 1995). The intrinsic
radio quasar luminosity function was assumed to be a power-law
$\Phi(L_R)\propto L_R^{-\beta}$ with arbitrary normalization and
slope $\beta=-3.0$, which is consistent with the bright-end slopes
measured for the optical and radio quasar luminosity functions
(e.g., De Zotti et al. 2005, Richards et al. 2006, Jiang et al.
2006)\footnote{The exact value of this slope does not affect our
results as the uncertainties in this parameter are degenerate with
the uncertainties in other model parameters}. We then assume that
the population of radio quasars is composed of three
sub-populations, as sketched in the right panel of
Figure~\ref{fig|cartoon}. The first one is represented by radio
BALQSOs, which are preferentially viewed at lines of sight close to
the disk or the torus, at a maximum angle of $\theta\lesssim
11.5^{\circ}$, corresponding to a fraction of $\sin(\theta)=20\%$,
as inferred in \S~\ref{sec|Results} and by DSS. Following Urry \&
Padovani (1995) we then assumed that the fraction of beamed sources
is viewed at an angle of $38^{\circ}$ from the jet axis,
corresponding to a fraction of $\sim 21\%$. The last sub-group is
composed of radio non-beamed sources viewed at intermediate angles
accounting for the remaining $\sim$ 59\% of the total. AI-BALQSOs in
this model must be randomly distributed at all angles to ensure
their fraction to be flat with radio power. The relative
contributions of the different type of sources to the total radio
luminosity function are shown in the left panel of
Figure~\ref{fig|simul}, plotted as $L^3\Phi(L)$ which flattens out
the luminosity function and enhances the difference among the lines.

The luminosity function of beamed sources was computed following the
method outlined in Urry \& Padovani (1991). The observed luminosity
function at a given observed luminosity $L_{\rm obs}$ was obtained
by convolving the intrinsic luminosity function with the probability
$P(L_{\rm obs}|L_{\rm int})$. Following Urry \& Padovani (1991) we
allowed beamed sources to also have a diffuse component from the
lobes and the core, in addition to the beamed emission. Therefore
the observed luminosity of this sub-group was parameterized as
$L_{\rm obs}=(1+f\delta^p)L_{\rm int}$, where $L_{\rm int}$ is the
intrinsic luminosity, $f$ is the fraction of luminosity in the jets,
$\delta=[\gamma(1-\beta \cos \theta)]^{-1}$ is the Doppler factor
corresponding to a given angle $\theta$ and Lorentz factor
$\gamma=(1-\beta^2)^{-1/2}$, and the exponent $p$ is a parameter
depending on the spectrum and reacceleration of the radiating
particles. Given the relatively high number of parameters in this
model, we adopt $p=3$, $\gamma=11$, $f=0.005$ and the maximum
beaming angle to $\theta_{\rm max}=38^{\circ}$. These are the same
values used by Urry \& Padovani (1995) and, more recently, by
Padovani et al. (2007) to reproduce within unification models the
luminosity distributions of samples dominated by beamed sources.

The apparent luminosity function of beamed radio sources,
``crosses'' the intrinsic luminosity function at a luminosity which
depends on the minimum luminosity in which we allow beamed sources
to appear in the luminosity function (see Figure~\ref{fig|simul}).
We parameterized this luminosity as $\log L_{1.4 {\rm \, GHz \, ,
MIN}}=32-\Delta \log L_{1.4 {\rm \, GHz}}$, where $\log L_{1.4 {\rm
\, GHz}}\sim 32$ is the minimum luminosity in the sample. The solid
line in the right panel of Figure~\ref{fig|simul} shows our best-fit
model for the expected fraction of BALQSOs as a function of radio
luminosity. Optimizing the model to the $L_{1.4 {\rm \, GHz
}}$-dependence of $f_{\rm BAL}$, we find $\Delta \log L_{1.4 {\rm \,
GHz }}=0.09^{+0.02}_{-0.01}$, with $\chi_{\rm min}^2\sim 12.3$, for
12 degrees of freedom. This model implies that beamed sources are
almost absent in the luminosity function below $L_{1.4 {\rm \, GHz
}}\sim 10^{32}\, {\rm erg\, s^{-1}}$ and then start to become
important only above this luminosity threshold. To better compare
with the data, we also show with open circles the expected fraction
for this model integrated over the same radio luminosity bins of the
data.

In this model the dependence of the BALQSO fraction on radio
luminosity is entirely due to geometry and it is independent of the
duty cycle of RQs within the overall quasar population.
Nevertheless, the origin of the radio emission can still be an
evolutionary phase or related to specific properties of quasars such
as the magnetic field, the black hole mass, spin or Eddington ratio.

A pure geometrical model, although plausible on statistical grounds,
is not entirely satisfactory. Polar BI-BALQSO outflows have indeed
been observed in several cases (e.g., Zhou et al. 2006; Ghosh \&
Punsly 2007; Wang et al. 2008), and the paucity of FRIIs within
BALQSOs can hardly be accounted for in simple orientation models
(see \S~\ref{subsec|evolutionmodel}). Nevertheless, we expect that
some non-orientation processes may effect the physics of radio and
BALQSOs and induce different evolutionary trends in different type
of sources. For example, it has been observed that AGNs have broad
Eddington ratio distributions at fixed black hole mass (e.g., Shen
et al. 2008). The more luminous FRII sources may be driven by higher
Eddington ratios than FRI sources, at fixed BH mass, and
consequently may have higher kinetic powers and stronger radiation
pressures. These may be more efficient in removing BAL clouds along
the line of sight and also be more efficient in driving collimated
jets, thus explaining both the reduced BAL fraction and the large
scale hot-spots observed at kpc scales (as also partly discussed by
Gregg et al. 2006).


\subsection{The Energy Exchange Model}
\label{subsec|energyexchange}

Another possible explanation for the drop in $f_{\rm BAL}$, could
rely on some sort of ``energy exchange'' from the wind to the jet in
high luminosity radio sources. Within the framework of the
geometrical model, the energy gained by the system from the external
gas accretion must be redistributed between the jet and the wind by
some physical mechanism, such as the magnetic field. We could then
expect that those quasars with high radio powers preferentially
inject energy into the jet due to a stronger intrinsic magnetic
field and/or a rapidly spinning central black hole, thus limiting
the energy channeled into the wind and inevitably causing a drop in
$f_{\rm BAL}$.

\begin{figure*}[ht!]
\epsscale{1.0}
\plotone{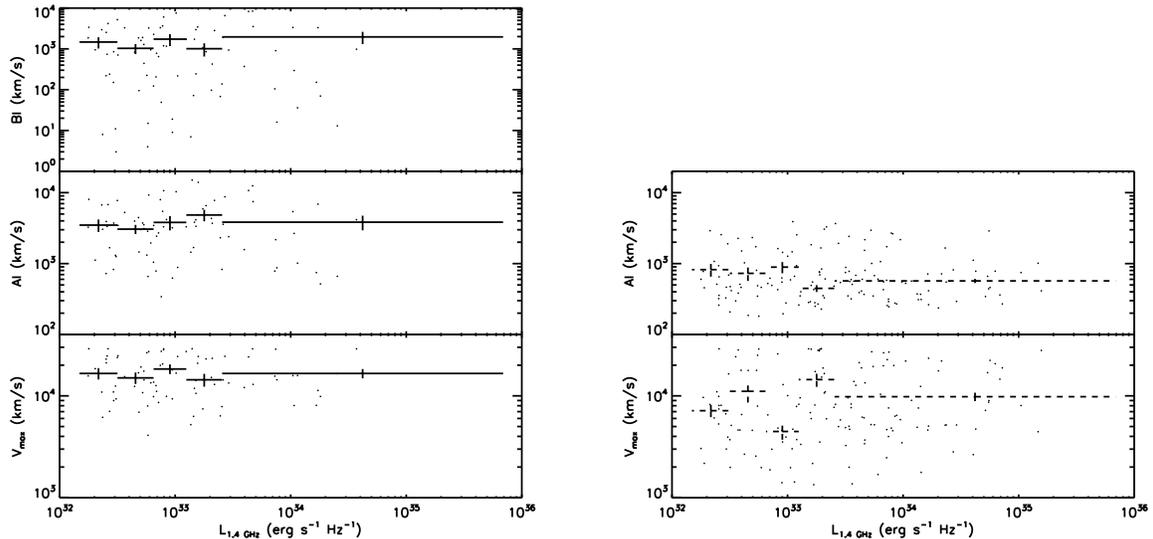} \caption{\emph{Left panel}: the \emph{crosses}
in the \emph{upper} panel show the mean BALQSO Balnicity Index per
bin of radio luminosity, while the full sample is shown with points;
the \emph{middle} panel shows the same for the Absorption Index,
while the \emph{lower} panel shows the maximum velocity measured
from the absorption troughs as a function of radio luminosity.
\emph{Right panel}: Absorption Index (\emph{upper} panel) and
maximum velocity (\emph{lower} panel) for AI but not BI-selected
BALSQSOs. We do not find any significant trend with radio luminosity
for all plotted quantities.} \label{fig|AI}
\end{figure*}

We tested this idea by considering in the left panel of
Figure~\ref{fig|AI} the BALQSO BI index (upper panel), AI index
(middle panel) and maximum velocity $V_{\rm max}$ (lower panel)
measured from the absorption troughs as proxy of the wind power, and
studied their behavior as a function of radio luminosity. In each
bin, the quantities were computed from the biweight-mean (Hoaglin et
al.\ 1983) and the errors on this mean were estimated from the
biweight-$\sigma$ reduced by $\sqrt{N}$, where $N$ is the number of
BALQSOs in the bin. Becker et al. (2000, 2001) found the mean BI
decreases with radio luminosity for the high-ionization BALQSOs
(HIBALQSOs), an effect which could be simply interpreted as a
decrease in the amount of obscuring mass for the more luminous
sources. They also found tentative evidence for HIBALQSOs to have
their maximum velocity decreasing with radio power. In our larger
sample, which should be dominated by HIBALQSOs, we instead find the
AI, BI and maximum wind velocity show no clear evidence for a
decrease at high radio powers although with a large scatter, in
conflict with the energy exchange model. We also find marginal
evidence for a drop in the AI index but not in the $V_{\rm max}$ of
the AI but not BI-defined BALQSOs (right panel of
Figure~\ref{fig|AI}).

The energy exchange model may also conflict with evidence that AGNs
have similar radiative and kinetic efficiencies (e.g., Shankar et
al. 2008a,b). Moreover, if the efficiency is constant, then $V_{\rm
max}$ should increase with luminosity (i.e., with the mean accretion
rate), rendering the result on its flatness rather intriguing (e.g.,
Ganguly et al. 2007). 

\section{DISCUSSIONS}
\label{sec|discu}

Synchrotron self-absorption may impose an uncertainty in estimating
the radio flux density. The spectral analysis by Becker et al.
(2000) found that about one third of the sources could show
synchrotron self-absorption at lower frequencies, typical of compact
radio sources. If so, up to 30\% of the radio luminosities of
BALQSOs may be underestimated and in turn significantly affect our
results. To quantify the implications of self-absorption, one could
assume the intrinsic fraction of BALQSOs to be $f_{\rm BAL}=k$, with
$k$ constant with increasing radio luminosity. If self-absorption
mainly affects sources above an intrinsic radio luminosity of
$L_{\rm break}=L_{1.4 {\rm \, GHz }}\sim 2\times 10^{33}\, {\rm
erg\, s^{-1}}$, as suggested by the observed behavior in $f_{\rm
BAL}$, then the \emph{observed} fraction of BALQSOs, $f_{\rm BAL\,
obs}$, will be ``distorted'' by the fact that a fraction of $\sim
30\%$ of luminous radio sources will be transferred from high to low
luminosities, decreasing the number of BALQSOs at high radio powers
and proportionally increasing it at low radio powers. For example,
by setting $k=0.15$, the fraction of BALQSOs with $L_{1.4 {\rm \,
GHz }}\gtrsim L_{\rm break}$ will be $f_{\rm BAL\,
obs}=[k\times(1-0.3)]/[k\times(1-0.3)+(1-k)]\sim 0.1$, while the
fraction of sources with $L_{1.4 {\rm \, GHz }}\lesssim L_{\rm
break}$ will be $f_{\rm BAL\,
obs}=[k\times(1+0.3)]/[k\times(1+0.3)+(1-k)]\sim 0.19$, which is
only in marginal agreement with observations. This simple
explanation therefore cannot fully account for the steep drop in
$f_{\rm BAL\, obs}$. To be relevant in our study, synchrotron
self-absorption should also be able to decrease the intrinsic radio
power by factors of 10 to 1000, however typical reductions of only a
factor of $\sim 10$ are expected (e.g., Snellen \& Schilizzi 1999).

\section{CONCLUSIONS}
\label{sec|conclu}

We find that the fraction of classical BALQSOs among the FIRST radio
sources in the SDSS DR3 catalogue, is $20.5^{+7.3}_{-5.9}$\% at the
faintest radio powers detected ($L_{\rm 1.4 GHz}\sim 10^{32}\, {\rm
erg\, s^{-1}}$), and rapidly drops to $\lesssim 8$\% at $L_{\rm
1.4\, GHz}\sim 3\times 10^{33}\, {\rm erg\, s^{-1}}$. Similarly,
adopting the broader AI definition of Trump et al. (2006) we find
the fraction of radio BALQSOs to be $44^{+8.1}_{-7.8}$\% reducing to
$23.1^{+7.3}_{-6.1}$ at high luminosities.

The detected fraction at low radio luminosities is consistent with
the recent estimates inferred in infrared bands by DSS, supporting
the fact that longer, less absorbed wavelengths, are more suitable
for probing the intrinsic fraction of BALQSOs. We therefore agree
with Hewett and Foltz (2003) who claimed a similar occurrence of
BALQSOs in optical and radio samples. The variation of the BALQSO
fraction with radio power is also similar to that found by Becker et
al. (2000, 2001) and Hewett and Foltz (2003), although our results
are supported by a much larger sample. The decrease in the number of
radio BALQSOs is real, in the sense that it is not dependent on
redshift or luminosity cuts, biased by optical or radio selection
effects, or dependent on the specific type of radio source
considered (core-dominated or extended with multiple components). We
also find that at fixed optical magnitude, the highest bins of radio
luminosity are preferentially populated by non-BALQSOs, consistent
with the overall trend.

However, we find that those quasars identified as AI-BALQSOs but
\emph{not} under the classical BI definition, do not show the same
significant drop in the fraction as a function of radio power and
share similar cumulative radio distributions as non-BALQSOs. This
further supports independent claims for which these sources
characterized by lower equivalent width, may represent an
independent class with respect to the classical BALQSOs. We find the
BI, AI and mean maximum wind velocity to be roughly constant at all
radio powers.

We discuss several plausible physical models which may explain the
observed fast drop in the fraction of the classical BALQSOs with
increasing radio power, \emph{although no one model is entirely
satisfactory}. A strictly evolutionary model for the BALQSO and
radio emission phases requires a strong fine-tuning to work, while a
simple geometric model where the apparent radio luminosity function
is partly due to beamed, non-BALQSOs, although more promising in
matching the data, cannot explain strong polar BALQSOs and the
paucity of FRII sources within BALQSOs. Of course, the drop in
$f_{\rm BAL}$ could also be viewed as an \emph{increase} of the
BALQSO fraction towards low-intermediate radio sources which may be
more likely to show BALQSO features. This model however does not
explain why the fraction of BALQSOs at low radio powers is identical
to the one obtained in the optical-NIR.

\acknowledgements We thank the anonymous referee for suggestions
which improved the paper. We also thank Chris Kochanek for helpful
comments that improved the presentation of this work, and Michael S.
Brotherton and Rajib Ganguly for several useful comments and
suggestions. We finally acknowledge discussion at the AGN lunch in
the OSU astronomy department.


\newpage

\newpage

\end{document}